\documentclass{elsart}
\usepackage{epsfig}

\begin{document}
  \runauthor{Piran, GRBs - A puzzle being resolved}
 \begin{frontmatter}

 \title{GAMMA-RAY BURSTS - A PUZZLE BEING RESOLVED}
\author{Tsvi Piran\thanksref{support}}
\address{Racah Institute for Physics, The Hebrew University, Jerusalem,  91904, Israel \thanksref{permanent}}
\address{Physics Department,  New York University, New York, NY 10003, USA}
\address{and}
\address{Physics Department, Columbia University, New York, NY 10027, USA}
\thanks[support]{Supported by the US-Israel BSF grant 95-328
and by NASA grant NAG5-3516.}
\thanks[permanent]{permanent address}

  \vspace*{0.9cm}
\begin{abstract}
{For a few seconds a gamma-ray burst (GRB) becomes the brightest
object in the Universe, over-shining the rest of the Universe
combined! Clearly this reflects extreme conditions that are
fascinating and worth exploring.  The recent discovery of GRB
afterglow have demonstrated that we are on the right track towards the
resolution of this long standing puzzle.  These observations have
confirmed the relativistic fireball model (more specifically the
internal-external shocks model).  The prompt optical emission seen in
GRB 990123 have demonstrated that GRBs involve ultra-relativistic
motion.  The breaks in the light curves of GRB 990123 and GRB 990510
and the peculiar light curves of GRB 980519 and GRB 980326 disclosed
that these GRBs are beamed.  I examine these recent developments and
discuss their implications to the models of the source.  I argue that
the current understanding implies that GRBs signal the birth of
stellar mass black holes.}
\end{abstract}   
\end{frontmatter}

\section{Prologue}
It is a well kept secret that Dave Schramm wrote a paper on GRBs.
Dave visited Jerusalem for the Supernovae winter school of 1989.  Our
discussions, together with Mario Livio and David Eichler, led to:
``Nucleosynthesis, Neutrino Bursts and Gamma-Rays from Coalescing
Neutron Stars" \cite{Eichler89} that was published in Nature in May
1989. At that time it was well known that binary neutron star mergers
are excellent sources of a characteristic gravitational radiation
signal as well as a strong but undetectable MeV neutrino burst.
Although the idea that cosmological neutron star mergers are sources
of GRBs was mentioned in passing in some earlier works, this was the
first systematic attempt to construct a ``first principles'' model
\cite{Nemiroff93}.  Additionally, we have  pointed out 
in this paper that these mergers could be important sites for
r-process nucleosynthesis.

Dave was very excited about this paper.  He predicted that it will be
widely quoted as it was at the intersection of four different topics:
Gravitational radiation, Neutrinos, GRBs and Nucleosynthesis, and all
four communities would refer to it. He was wrong, at least for a
while. At first, the paper was ignored by all four communities.  In
1989 GRBs were believed to be galactic and anything associated with a
cosmological GRB model was too controversial to be taken
seriously. Cosmological neutron star mergers were discussed only as
sources of gravitational radiation.  This has changed gradually after
BATSE's discovery in 1991 that GRBs are cosmological \cite{Meegan92}.
Eventually neutron star mergers became the ``canonical'' GRB model,
and the paper begun to be cited.  Moreover, recent observations
\cite{Cowan99} and detailed calculations
\cite{Freiburghaus99} suggest that neutron star mergers are indeed 
important sites for r process nucleosynthesis. Even among the
gravitational radiation community there seems to be a beginning of a
discussion of the implication of a GRB association for the detection of
gravitational waves \cite{Finn99}.  Like in many other issue Dave was
right. It just took the rest of us longer to realize that.

  \section{Introduction}

Gamma-Ray Bursts (GRBs) are short and intense bursts of MeV range
$\gamma$-rays. During the last decade observational progress has
revolutionized our understanding of GRBs. BATSE on Compton-GRO have
found that GRBs are distributed isotropically revealing their
cosmological origin. More recently BeppoSAX discovered X-ray afterglow
\cite{Costa97}. This enabled accurate position determination and the
discovery of optical \cite{Paradijs97} and radio \cite{Frail97a}
afterglows and  host galaxies. Remarkably the afterglow
is a simple phenomenon that can be analyzed using a rather simple
model. The resulting information tells us a lot about the properties
of the GRB.

In this review I summarize the recent developments in GRB physics.  It
can be views as an update of my recent Physics Reports review
\cite{Piran99}. My goal, here, is to confront the predictions of the
Relativistic Fireball model with recent observations.  This model
had recently several successes in predicting the afterglow, the prompt
optical flash and the beaming break in the light curve.  The fireball
model deals with the ``outer'' radiating regions. It doesn't deal
directly with the ``inner engine'' - the source of the relativistic
ejecta that powers the whole phenomenon.  I review the implications of 
the current observations, as interpreted within the fireball model, to
models of the source and I discuss some of the possible sources. I
examine some open questions and in particular the possible ``energy crisis''.
I conclude with a ``wish list'' of new observations that might resolve
this enigma.

\section{Summary of GRB Observations}

\begin{figure}
\centerline{\epsfig{file=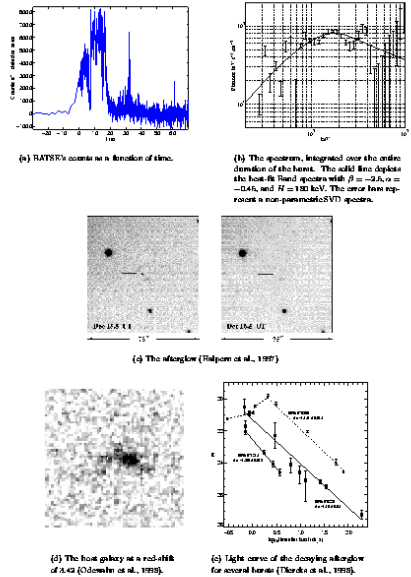,width=8cm}}
\caption{GRB 971214: Top left gamma-ray light curve; Top right the
spectrum; Middle the afterglow \cite{Halpernetal97}; Bottom left
the host galaxy \cite{Odewahn98}; 
Bottom right optical
light curve for GRB 971214, GRB 970508 and GRB 970228 \cite{Diercks98}.}
\label{GRB971214}
\end{figure}

GRBs are short and intense bursts of 100keV - a few MeV gamma-rays.
GRBs vary greatly from one to another.  Even though there is no
``typical'' GRB, I show in Fig. \ref{GRB971214} some features of GRB
971214 as an indication of what is a GRB.  In the following I
summarize briefly some observational facts concerning GRBs. I focus on
those observations that provide the basic clues for the theoretical
modeling.

\begin{itemize}
\item 
The observed fluence on earth is $10^{-5}-10^{-7}$ ergs/cm$^2$ (the
upper limit depends, of course, on the duration of the observations
while the lower limit depends, of course, on the detector).  Redshift
measurements of half a dozen GRBs indicate that for isotropic emission
the total energy is of the order $10^{51}-10^{54}$ergs.  Recent
observations suggest that the radiation is beamed, lowering this
energy by a factor of several hundred.

\item
The GRB spectrum is nonthermal. In most cases there is a strong power
law high energy tail extending to very high energies (up to a few GeV).
This nonthermal spectrum, which confused researchers at first,
provided the first and the most important clue to the nature of GRBs.
The spectrum can be fitted by the schematic Band
function \cite{Band93}: two power law joined smoothly together. The
lower energy spectral index, $d N(E)/dE$, is, quite generally, in the
range (1/3, -1/2) which is compatible with the relativistic synchrotron
model \cite{Cohen_etal_97}.  However, there are
indications \cite{Preece98} that in some bursts the lower energy slope
is steeper, contradicting this model.

\item
The duration ranges from a fraction of a second to several hundred
seconds. The light curve is extremely irregular and large fluctuations
on a time scale of up to the detection limit of msec are observed.
This high variability provides a second important clue to the
nature of GRBs.

\item
There are indications that shorter GRBs compose a different
subgroup. It is more difficult to detect short bursts and the observed
short bursts are, therefore, nearer than longer ones \cite{MNP94}. As a
result the short bursts distribution is compatible with a homogeneous
Eucleadian distribution \cite{Cohen_Piran95,KatzCanel96} in which $d\ln
N/d\ln S = -{3/2}$. There have been claims that long bursts with no
high energy component (no emission in the higher BATSE channels) are
also distributed with a $-{3/2}$ slope \cite{Tavani98}. However, it is
not clear if there is a real need to consider those as a separate
subgroup \cite{BonnellNorris99}.

\end{itemize}

\section{Afterglow Observations}

The Italian Dutch satellite, BeppoSAX, discovered X-ray afterglow on
February 28th 1997 \cite{Costa97}. By now X-ray afterglow has been
observed from two dozen bursts. As BeppoSAX can trigger only on long
bursts it is not known if short bursts are also followed by
afterglow\footnote{Hopefully, the planed mission of HETE II and the
proposed missions SWIFT and BALLERINA will answer this
question.}. Optical \cite{Paradijs97} and radio \cite{Frail97a}
afterglows were discovered using the accurate position obtained by
BeppoSAX in about half of the cases in which X-ray afterglow was
seen. It is not clear what determines whether optical and radio
afterglow are observed.

\begin{itemize} 
\item 
The energy involved can be estimated from the late phases of the optical
 \cite{Galamaetal98a,Granot99,Wijers_Galama99,Vreeswijk99} and the radio \cite{Waxman99} afterglows.
The overall energy emitted in the afterglow is of order
$10^{50}-10^{52}$. Quite generally it is only a fraction 
of the energy emitted during the GRB.

\item 
The afterglow light curve decays, in most cases, as a single power law
in time\footnote{Sari, Piran and Narayan \cite{SaPN98} have introduced
in the astro-ph version the notation $F_\nu \propto
t^{-\alpha}\nu^\beta$.  This notation was changed to
$t^{-\beta}\nu^\alpha$ in the Ap. J. version and in my
review \cite{Piran99}.  However, the astro-ph notation caught so well
that I return to this original notation here and elsewhere.}: $F_\nu
\propto t^{-\alpha}$, with $\alpha \sim 1.2$.  In two cases (GRB
980326 and GRB 980519) $\alpha \sim 2$ and in two cases GRB 990123 and
GRB 990510 there is a clear break from a flat decay $\alpha \sim
1.1-1.2$ to a much faster decline.  We show later that
these features indicate that these GRBs are narrowly beamed (we also 
clarify in that section the confusion between relativistic and geometric
beaming).

\item Prompt optical emission was observed from GRB 990123. This emission 
peaked with a ninth magnitude signal which lagged 70 seconds after the
gamma-ray peak and coincided with the prompt X-ray peak.
\end{itemize}

\section{The Relativistic Internal-External Shocks Fireball Model}

The nonthermal spectrum indicates that the observed emission emerges
from an optically thin region. However, a simple estimate of the
number of photons above $500$keV and a simple estimate of 
the size of the source ($c \delta t$
as implied by the observed variability) shows that such a  source must be
extremely optically thick to pair creation
\cite{Piran_Shemi,Fenimore93,WoodLoeb95,Piran97}. Such a source cannot
emit nonthermal emission. This is the Compactness
problem.

The simplest way to overcome this problem is if the source is moving
ultra-relativistically towards us \cite{Ruderman75,Goo86,KroPie}.
This has lead to the relativistic fireball model.  According to this
model slowing relativistic ejecta produces the GRBs and their
afterglow. Relativistic electrons that have been accelerate in the
relativistic shocks that form, emit the observed gamma-rays via
synchrotron or synchrotron self-Compton emission.  Both the energy
density of these electrons and of the magnetic fields should be close
to equipartition for efficient emission. To bypass causality and the
compactness limits the shocks must be extremely relativistic with
$\gamma \ge 100$.

This suggests that GRBs involve three stages:
\begin{itemize}
\item First a source produces a relativistic energy flow.
The observed fluctuations in the GRB light curves and the huge energy
released  indicates that the source is compact.  This ``inner engine''
is  hidden and it is not observed directly. This makes it difficult
to constrain GRB models and leaves only circumstantial evidence on the
nature of the sources.

\item The energy is transfered relativistically from 
the compact source to distances larger than $\sim 10^{13}$cm where the
system is optically thin. This is most likely in the form of a
relativistic particles flow but the possibility of a Poynting Flux
should also be considered. As we discuss shortly the flow must be
highly irregular to produce internal shocks.

\item 
The relativistic ejecta is slowed down and the shocks that form
convert the kinetic energy to internal energy of accelerated
particles, which in turn emit the observed gamma-rays.
\end{itemize}

External shocks arise due to the interaction of the relativistic
matter with the surrounding matter \cite{MR1}, like the ISM or with a
circumstellar wind that took place during an earlier epoch.  These
shocks are the relativistic analogues of SNRs.  Like in SNRs these
shocks are collisionless.  External shocks become effective at
\cite{SaP95} ${\rm min}(R_\gamma,R_\Delta)$, where $R_\gamma
\equiv l/\gamma_0^{2/3}$ and $R_\Delta \equiv l^{3/4}
\Delta^{1/4}$. 
$\gamma$ and $\Delta$  are the  Lorentz factor and the 
width of the shell (in the observer frame) and the subscript 0
indicated, here and elsewhere the initial value. The Sedov length, $l
\equiv \big(E_0 /[ (4 \pi /3) n_{ism} m_p c^2]\big)^{1/3}$, is the radius within
which the rest mass energy of the external material, whose density is
$n_{ism}$ equals the initial energy of the ejecta, $E_0$. Typical
values are: $l \sim 10^{18}$cm and $R_\gamma
\sim R_\Delta \sim 10^{15}-10^{16}$cm.

\begin{figure}
\centerline{\epsfig{file=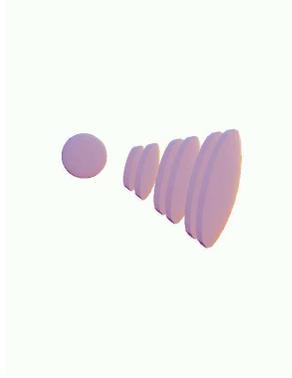,height=5cm}}
\caption{Internal shocks:  Faster shells catch up with slower ones 
and collide, converting some of their kinetic energy to internal
energy.}
\label{internal}
\end{figure}

\begin{figure}
\centerline{\epsfig{file=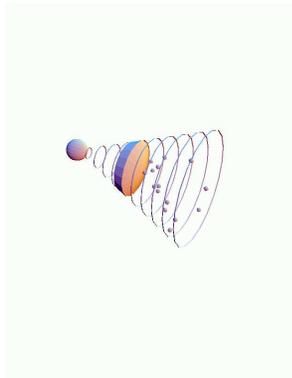,height=5cm}}
\caption{To produce variability in external shocks
the relativistic ejecta must encounters interstellar bubbles.
However, most of the shell passes between the bubbles and its kinetic
energy is lost.}
\label{external}
\end{figure}

Sari and Piran \cite{SaP97a} and Fenimore, Madras and
Nayakshin \cite{FenMadNaya} have shown that that external shocks cannot
produce efficiently the observed highly variable temporal structure
seen in GRB\footnote{Dermer and Mitman  \cite{Dermer99} suggest that
within external shocks with a very inhomogeneous media (see
Fig. \ref{external}), subpulses close to the line of sight could
produce the observed peaks.  But these subpulses cannot produce the large
variability in amplitude observed. Additionally, there is no
explanation why the expected correlation between arrival time and
subpulses duration is not seen.}.  Thus, GRBs are produce by the only
other alternative, internal shocks \cite{NPP92,PacXu,MR4}.  These
shocks arise in irregular flow when faster shells overtake slower ones
(see Fig. \ref{internal}). If the flow varies on a scale length
$\delta$ then internal shocks would take place at $R_{int} \sim
\delta \gamma_0^2$. $\delta = c \delta t $ can be inferred from the
observed temporal variability $ \delta t \le 1$sec indicating that
these shocks take place at $\sim 10^{13}$cm. If the shocks arise
earlier the system is still optically thick and the radiation does not
escape.  The observed GRB time scales reflect the time scales of the
``inner engine'' \cite{KPS97}.  The GRB duration corresponds to the
time that the ``inner engine'' is active.

Within the external shocks model late subpulses, which are produced
when the Lorentz factor is lower and at larger radii, are expected to
be longer.  The lack of any correlation between the width of subpulses
and their time of arrival is inconsistent with this
model \cite{Fenimore98}.  The lack of a direct scaling between the GRB
and the afterglow is another evidence for the internal shocks model.
This is a very important clue on the nature of the source. External
shocks can be produced by an explosive event in which the ``inner
engine'' releases all its energy at once.  The observed temporal
structure could have been produced in this case within the shocks due
to irregularities in the surrounding material  (see
Fig. \ref{external}).  However this would have been extremely
inefficient \cite{SaP97a}.  This rules out the possibility of an
explosive ``inner engine''.

Internal shocks can extract only a fraction of the total
energy \cite{MMM95,KPS97,Daigne98}.  Sari and Piran \cite{SaP97a}
suggested that the remaining energy will be extracted later via
external shock giving rise to additional emission at different
wavelengths - the afterglow.  Thus, GRB have a forth stage:
\begin{itemize}
\item 
The relativistic flow, which have been slowed down, but has not been
stopped, is slowed further by the surrounding material producing the
afterglow. This phase is regular and it can be modeled rather well by
the adiabatic Blandford-McKee \cite{BM} adiabatic self-similar
solution.
\end{itemize}

\section{The Late Afterglow: A Prediction and a Confirmation of the Fireball Model}

Paczy\'nski and Rhoads \cite{PacRho93} and independently
Katz \cite{Katz94} realized early on that the long term interaction of
the relativistic ejecta with the surrounding matter will produce a low
frequency afterglow. Paczy\'nski and Rhoads \cite{PacRho93} discuss
radio afterglow. Katz \cite{Katz94} discusses optical afterglow.
Later, M\'esz\'aros and Rees \cite{MR97} and independently
Vietri \cite{Vietri97} developed detailed models of this afterglow.  In
all these studies, it was thought that both the GRB and the afterglow
are produced by external shocks. This afterglow would have been the
long time extrapolation of the GRB and it should scale with the GRB
properties.

If GRBs arises due to internal shocks and the afterglow is produced by
external shocks \cite{SaP97a} we don't expect
direct scaling between the two.  In this ``Internal-External'' model
the GRB and the afterglow are produced by two different processes.
The recent afterglow observation provide a significant evidence for this
picture.


The afterglow provided additional direct confirmation of the fireball
model. The radio afterglow of GRB 970508 showed significant flickering
during the first few weeks.  This flickering decreased and eventually
stopped after about a month. Goodman \cite{Goodman97} quickly suggested
that the flickering arises arose due to scintillation. Initially the
source is small and it is within the scintillation regime. As the
source expands the scintillation stop. Using this idea Frail
et. al \cite{Frail97a} estimated that the size of the afterglow of GRB
970508 was $\sim 10^{17}$cm one month after the burst. Even before GRB
970508 Katz and Piran \cite{KatzPiran97} suggested that synchrotron
self-absorption would result in a rising spectrum in radio frequencies
for which the source is optically thick and using the observed flux and
an estimate of the temperature of the emitting regions one could
estimate the size of the emitting region.  Frail
et. al., \cite{Frail97a} obtain a size of $\sim 10^{17}$cm after one
month.  The agreement between the two independent estimates is
reassuring. These observations imply that the fireball is expanding
relativistically, and provided, for the first time a confirmation of
the notion of relativistic motion in GRBs.

\section{Synchrotron spectrum and Afterglow Observations}

The generic emission process for both the GRB and the afterglow is
synchrotron. It is generally assumed that the emitting electrons have
a power law energy distribution: $N(E) \propto E^{-p}$.  A typical
value that fits both the GRB and the afterglow observations is $p \sim
2.5$ \footnote{One can expect that this number will be universal and
won't vary from one burst to another.}. With $p=2.5$ the distribution
diverges at low energies, and there must be a low energy cutoff, which
is determined by the energy density available: $E_{min} = ???
e_{e}/[(p-2)n_e]$, where $n_e$ and $e_e$ are the electrons' density and
their energy density. The largest number of electrons is around $E_{min}$
and hence this is also the characteristic electron energy.  We denote
by $\nu_m$ the synchrotron frequency of an electron with this
energy. This is the ``typical'' synchrotron frequency.  The electrons'
energy density as well as the magnetic field energy density are
characterized as fractions $\epsilon_e$ and $\epsilon_B$ of the total
internal energy \cite{SaNP96}.

These definitions enable Sari, Piran and Narayan \cite{SaPN98} to estimate
the instantaneous spectrum: 
The low (but not extremely low) frequency spectrum is
given  by the low energy synchrotron tail: $F_\nu \propto
\nu^{-1/3}$. At intermediate frequencies the spectrum depends on the
whether the ``typical'' electrons are cooling within the hydrodynamic
time $t_{hyd}$.  In a fast cooling system the cooling time of an 
electron with $E_{min}$
is shorter than $t_{hyd}$. We define the cooling frequency, $\nu_c$,
as the synchrotron frequency of an electron that cools during the
local hydrodynamic time scale: $E_c/P_{\nu_c} (E_c) = t_{hyd}$. For
fast cooling $\nu_c < \nu_m$ and  $F_\nu \propto
\nu^{-1/2}$ for slow cooling $\nu_m < \nu_c$ and $F_\nu \propto
\nu^{-(p-1)/2}$.  The highest part of the spectrum is always dominated
by emission from fast cooling electrons with: $F_\nu \propto
\nu^{-p/2}$. At very low frequencies, usually at radio frequency, the
system may become optically thick to synchrotron self absorption. We
denote by $\nu_{sa}$ the self absorption frequency for which $\tau
(\nu_{sa}) = 1$.  Unlike the common discussion of synchrotron self
absorption in text books (which assumes $\nu_{sa} > \nu_m$ and obtain
$F_\nu \propto \nu^{5/2}$), here $\nu_{sa} \ll \nu_m$.  In this case
$F_{\nu_{sa}} \propto \nu^2$, just like the usual Wien part of a black body
spectrum.  Combined we have $F_\nu \propto \nu^{\beta}$ with :
\begin{equation}
\beta = \cases{ 2 & for $\nu < \nu_{sa}$ -  self absorption; \cr
          1/3 &  for $\nu_{sa}< \nu < {\rm min}(\nu_m,\nu_c)$; \cr
   - 1/2 &  for $\nu_c < \nu < \nu_m $  - fast cooling; \cr 
-(p-1)/2 & for  $\nu_m < \nu < \nu_c $  - slow cooling; \cr
             -p/2 & for  $ {\rm max}(\nu_m,\nu_c) <\nu $. \cr }
\end{equation}

\begin{figure}
\centerline{\epsfig{file=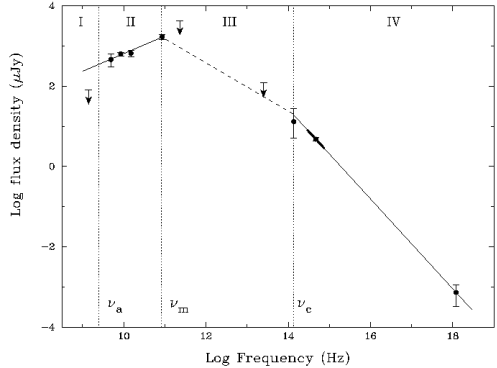,height=5cm}}
\caption{The X-ray to radio spectrum of GRB 970508 on May 21.0 UT (12.1
days after the event). The fit to the low-frequency part,
$\alpha_{\rm 4.86-86 GHz}$ = 0.44
$\pm$ 0.07, is shown as well as the extrapolation from X-ray to
optical (solid lines). The local optical slope (2.1--5.0 days after
the event) is indicated by the thick solid line. Also indicated is
the extrapolation $F_{\nu} \propto
\nu^{-0.6}$ (lines). Indicated are the rough estimates of the break
frequencies $\nu_{\rm a}$, $\nu_{\rm m}$ and $\nu_{\rm c}$ for May
21.0 UT from Galama et al., \cite{Galamaetal98a}.}
\label{spectrum}
\end{figure}

The resulting spectrum is a combination of four power laws, with three
of the four slopes fixed and one depending on whether the electron are
fast cooling or not.  A comparison of the theoretical model and the
observation is shown in Fig. \ref{spectrum} 
 \cite{Galamaetal98a}. The agreement is remarkable for such a
simple model.

In reality the emitting region is inhomogeneous and different parts are
moving with a different Lorentz factors.  A specific
example \cite{Granot99} of integration over an inhomogeneous
Blandford-McKee solution including different viewing angles results in
a spectrum which is basically similar to the one shown above but with
the sharp corner replaced by smooth curves.

This instantaneous spectrum is valid during the GRB phase and during
the later afterglow phase. However, the GRB phase involves
simultaneously emission from multiple shocks and the combined spectrum
might be more complicated.  We still expect generically that the low
frequency (say X-ray, for the GRB phase) slope will always be less
steep then 1/3. Cohen et al. \cite{Cohen_etal_97} find that this is
satisfied in several strong bursts that have a well determined
spectrum. Preece et. al. \cite{Preece98} have found that a few percent
of the bursts have a steeper low energy slope, suggesting that some
modification to the simple synchrotron model may be needed.  On the
other hand, at high frequency, one should consider the effects of
inverse Compton scattering which might dominate over synchrotron
cooling \cite{SaP97b}.  Ghisellini and Celotti \cite{Ghisellini99} even
suggest that numerous pairs are produces and their inverse Compton
emission dominate the observed gamma-ray emission during the GRB.

To determine the light curve we need $\nu_{sa}$, $\nu_m$, $\nu_c$, and
$F(\nu_m)$ (or $F_{\nu_c})$ as a function of time (all other fluxes
are determined by these quantities).  These depend, in turn, on the
energy density, $e$, the electron density, $n_e$, and on R and
$\gamma$.  Sari, Piran and Narayan \cite{SaPN98} used the scaling laws
of the Blandford-McKee solution (which assumes adiabatic evolution and
propagation into a constant density surrounding medium) 
and the scaling laws of a radiative solution to obtain 
specific light curves.  The two most important equations are the
adiabatic energy equation and the photon's arrival time (the
detector's time) equation\footnote{When using the full Blandford-McKee
solution there are factors of order unity on the l.h.s of these
equations \cite{Sari97a}.}:
\begin{equation}
E_0 = M(R) c^2 \gamma^2 ,
\label{adiabatic}
\end{equation}
and 
\begin{equation}
t_{obs} = {R \over 2 c \gamma^2} ,
\label{time}
\end{equation}
where $M(R)$ is the accumulated  mass at a radius $R$. Using these
equations we can express $R$ and $\gamma$ as a function of the
detector's time, and from this we can obtain the rest of the
equations.  The most important feature of the hydrodynamic solution is
that the radius is changing much slower with time than the Lorentz
factor. The system can almost be viewed as standing still with the
Lorentz factor decreasing in time. However, the weak dependence of $R$
on $t$ does make some difference as this is the main feature that
varies from one solution (e.g. adiabatic) to another (e.g. radiative).

A nice feature that arises is a simple relation between $\alpha$,
$\beta$ and $p$. For the  spherical adiabatic case we have:
\begin{equation}
\alpha =\cases { 3 \beta/2= 3(p-1)/4   & for $\nu < \nu_c$,
\cr (3 \beta-1)/2  = (3p-2)/4 & for $\nu > \nu_c$.}
\label{beta_spherical}
\end{equation}
These relations are satisfied, for example, for the afterglow of GRB
970508 for which $\alpha = 1.12$ and $\beta = 1.14$ corresponding to
$p \sim 2.4$.

\section{The Early Afterglow and the Prompt Optical Flash}

Radio flickering and the rising radio spectrum of the afterglow of GRB
970508 have shown that this afterglow was expanding relativistically
one month after the burst.  However, at this stage it has slowed down
significantly and its Lorentz factor was of order a few. This is not
the ultra-relativistic motion expected during the burst itself. Can we
determine what is the initial Lorentz factor?  This important question
could provide another clue to the nature of the relativistic flux and
could distinguish between different source models. An extremely high
Lorentz factor (of order $10^5$ or so) would indicate a Poynting
flux \cite{Usov92,Thompson94}.  While a lower Lorentz factor does not
rule out a Poynting flux, it is an indication in favor of a Baryonic
flow.

There are already some limits on the initial Lorentz factor.  The
Compactness problem provide a lower limit of $\sim 100$.  Modeling of
the internal shocks emission suggests an upper limit of 
$\sim 1000$\cite{SaP97b}. However, we would like a direct measurement. For this our
best bet is to turn to the very early afterglow - to the initial
stages of the interaction of the ejecta with the ISM.  This phase is
almost simultaneous with the GRB itself \cite{Sari97a}. Thus, its
detection poses an observational challenge in obtaining quickly an
accurate position and following it up with a rapid response.

Unlike the late adiabatic afterglow, the early afterglow is most
likely radiative.  Namely, the energy carried away by the emitted
radiation is significant compared to the total energy and these losses
influence the dynamics.  A radiative system must be fast cooling (if
the ``typical'' electron does not cool there cannot be efficient
energy losses).  To obtain the radiative light curve we replace the
adiabatic energy equation \ref{adiabatic} by the radiative energy
equation:
\begin{equation}
E_0 = M(R) c^2 \gamma (R) \gamma_0  .
\end{equation}
We also replace the emissivity equation.  For slow cooling the emitted
flux is: $N_e P$, where $N_e$ is the number of emitting electrons and
$P$ is the synchrotron emissivity. For fast cooling the emitted flux
is $dE_{hyd}/dt$: all the energy produced by the shock is radiated
away.

\begin{figure}
\centerline{\epsfig{file=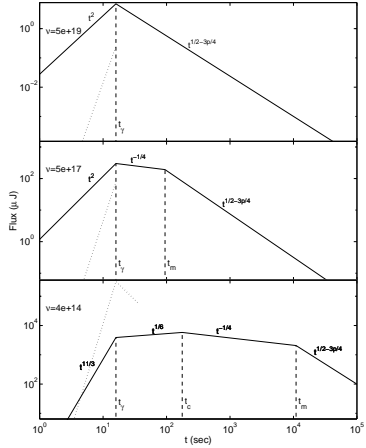,height=6cm}}
\caption{The early gamma-ray, X-ray and optical afterglow. The dashed
curve on the optical range is due to the reverse shock
emission \cite{SaP99a,SaP99b}. }
\label{early_afterglow}
\end{figure}

The early afterglow peaks in the soft gamma-rays or hard
X-rays \cite{SaP99b}. This emission could be viewed as a superposition
of a longer and smoother and somewhat softer component on top of the
variable, hard internal shocks signal \cite{Sari97a}. This signal
should be scaled and compared with the late afterglow as they arise
from  basically the same phenomenon.  Indeed in many cases
BeppoSAX's late time X-ray data  extrapolates nicely with a
$t^{-1.3}$ power law decay to the initial observations.

A remarkable new component that appears in Fig. \ref{early_afterglow}
is the prompt optical flash. Early Gamma-rays and X-rays and late
optical afterglow emission arises from the forward shock: i.e. from
shocked ISM material.  During the early interaction of the ejecta with
the ISM, before the solution has settled down to the Blandford-McKee
self similar solution, there is also a reverse shock that heats the
ejecta\footnote{This notation is somewhat confusing as this reverse
shock propagate backwards only in the fluid's frame. It is propagating
relativistically towards the observer in the observer frame.}.  The
emission from the reverse shock peaks at lower frequencies:
$\nu_{m_{rev}} \approx \nu_{m_{for}}/\gamma^2$.  If $\nu_{m_{for}}$ is
in the X-ray and $\gamma_0 \sim 100$ we expect $\nu_{m_{rev}}$ to be
in the optical or UV.  Comparable amounts of energy are generated by
the forward and the reverse shocks, and this energy is comparable to
the energy of the GRB itself.  A strong 5th magnitude optical flash
would have been produced if the fluence of a moderately strong GRBs,
$10^{-5}$erg/s/cm$^2$ would have been released on a time scale of
10sec in the optical band.  Even a small fraction of this would be
easily observed.

While the synchrotron frequency of the reverse shock is much lower
than the synchrotron frequency of the forward one, both have the same
cooling frequency (both have the same magnetic field and the same
bulk Lorentz factor).  As the forward shock must be radiative:
$\nu_{c_{for}} < \nu_{m_{for}}$. If $\nu_{c_{rev}}=\nu_{c_{for}} \gg
\nu_{m_{rev}}$ the reverse shock might not be radiating efficiently 
and this might lower the observed signal. Table \ref{t:afterglow} from
 \cite{SaP99b} show this reduction. Even with these factors,
one can hardly avoid a strong optical emission from the reverse shock.
Additional effects such  as inverse Compton scattering and self
absorption can somewhat reduce this flux, but even so a signal
stronger than 15th magnitude is expected.

\begin{table}
\begin{center}
\begin{tabular}{|c|c|c|}
\hline
& $\nu_c>\nu_{op}$ & $\nu_c<\nu_{op}$ \\ \hline\hline
$\nu_m>\nu_{op}$ & $\left( \frac {\min(\nu_c,\nu_m)} {\max(\nu_c,\nu_m)}
\right) ^{\frac 1 2} \left( \frac {\nu_{op}} {\min(\nu_c,\nu_m)} \right)^{\frac 4 3^{%
\phantom{1}}}_{\phantom{1_1}} $ & $\left( \frac {\nu_{op}} {\nu_m}
\right)^{1/2}$ \\ \hline
$\nu_m<\nu_{op}$ & $\left( \frac {\nu_{op}} {\nu_m} \right)^{-(p-2)/2}
\left( \frac {\nu_{op}} {\nu_c} \right)^{1/2}$ & $\left( \frac {\nu_{op}}
{\nu_m} \right)^{-\frac {p-2} 2 ^{\phantom{1}}} _{\phantom{1_1}}$ \\ \hline
\end{tabular}
\end{center}
\label{t:afterglow}
\caption{The fraction of the energy  emitted 
in the optical frequency $\nu_{op}$, as function of the cooling
frequency $\nu_c$ and the typical frequency $\nu_m$ \cite{SaP99b}.}
\end{table}

\begin{figure}
\centerline{\epsfig{file=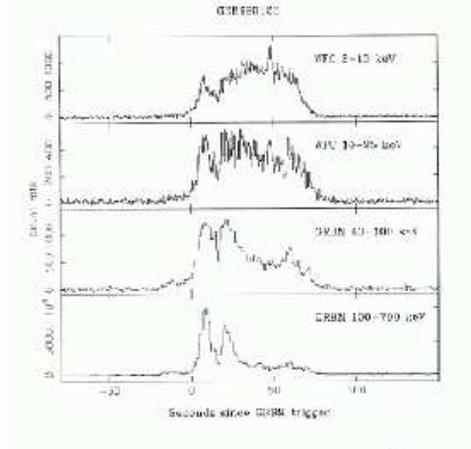,height=6cm}}
\caption{The light curves of  GRB 990123 in different 
bands \cite{Costa99}: from top to bottom: 2-13keV; 13-40keV; 40-100keV
and 100-700keV. The increasing X-ray flux at late time is seen
clearly.  Note that BeppoSAX's trigger is $\sim 25$sec after BATSE's
trigger. }
\label{990123_light}
\end{figure}

\begin{figure}
\centerline{\epsfig{file=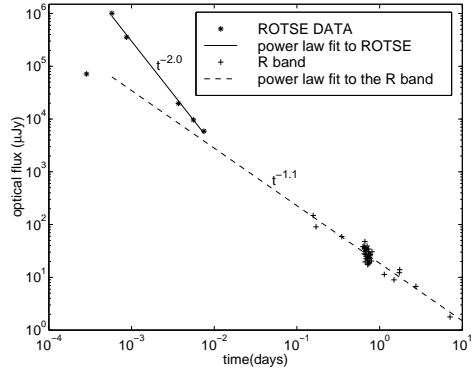,height=5cm}}
\caption{The early optical light curve of GRB 990123 \cite{SaP99c}. }
\label{990123_optical}
\end{figure}

Sari and Piran \cite{SaP99a} presented this detailed prediction of a
strong optical flash in the Rome meeting, that took place in October
1998.  The possibility of strong optical emission was noticed by
M\'esz\'aros \& Rees \cite{MR97} in two of several models they
examined.  This prediction was almost in contradiction with the LOTIS
upper limits on several bursts (see e.g. Williams et al. \cite{LOTIS}).
Less than two month later on January 23 1999 ROTSE \cite{Akerloff99}
was triggered by the GCN network \cite{Barthelmy94} and detected a 9th
magnitude optical signal accompanying GRB 990123. A careful
examination of the different light curves of GRB 990123 (see
Fig. \ref{990123_light}) reveals that the peak optical emission was
offset by at least 25 seconds from the peak gamma-ray emission. In
fact, there is no correlation between the gamma-rays and the optical
photons.  Moreover, the soft X-ray signal peaks at the time of the
optical peak and quite generally the spectrum evolves from hard to
soft. All this is in complete agreement with the picture presented
above. The early afterglow slightly lags after the GRB. The forward
shock emission peaks in X-rays while the reverse shock emission peaks
at the optical or UV band. The optical light curve shows an initial
phase of rapid decline $\sim t^{-2}$ again in a complete agreement
with the prediction of the reverse shock
emission \cite{SaP99a,SaP99b,KS99}. Later on this turns to the common
$t^{-1.1}$ decay seen in other bursts (see Fig. \ref{990123_optical}).

These observations provide us with three independent estimates of the
Lorentz factor during the early afterglow \cite{SaP99a}:
\begin{itemize}

\item The time of the optical flash peak, $\sim 70{\rm sec} \sim
l/ (c \gamma^{8/3})$. For $l \sim 10^{18}$cm equation \ref{time}
yields: $\gamma_0
\sim 200$.

\item The initial decay like $t^{-2}$ suggests that the typical
synchrotron frequency was below the optical early on. This suggests
$\gamma_0 \sim 200$.

\item The initial decline of the X-ray suggests that already initially the
typical synchrotron frequency was below the 1.5-10keV band.  Using the
initial ratio $\nu_{m_{for}}/\nu_{m_{rev}}$ we find that $\gamma_0
\sim 70$.

\end{itemize}

\begin{figure}
\centerline{\epsfig{file=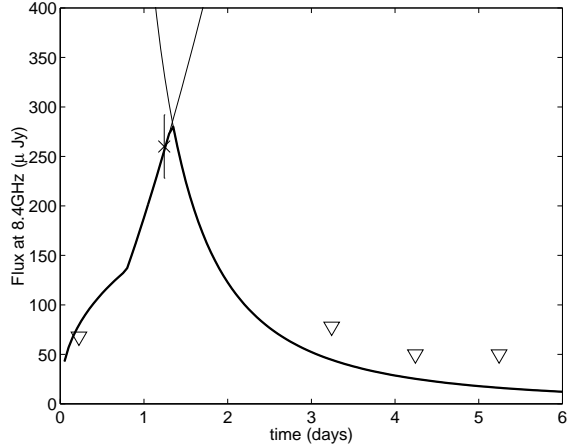,height=6cm}}
\caption{Observations and theoretical light curve for the  early radio 
signal of GRB 990123 \cite{SaP99c}. }
\label{990123-radio}
\end{figure}

Radio emission was observed from GRB
990123 \cite{Kulkarni99b,Galamaetal99}. This radio emission peaked
around one day with marginal detections prior and later than that
(see Fig. \ref{990123-radio}). This radio emission is also produced by
the reverse shock. The expected radio light curve (see
Fig. \ref{990123-radio}) was calculated using the optical
data \cite{SaP99c}. The fit to the observations is almost too nice. The
radio emission is suppressed early on due to synchrotron self
absorption. This enables us to estimate the perpendicular 
size of the system , $\gamma c t$, at
 one day after the burst to be $\sim 10^{15}$cm!

\section{Beaming, Jets and Flying Pancakes}
\label{jets}
If we assume isotropic emission then the energies emitted by GRB
971214, at z=3.418 and by GRB 990123 at z=1.65 are $\sim 3 \times
10^{53}$ergs and $4 \times 10^{54}$ergs respectively\footnote{This values
depend, of course, on the cosmological model assumed.}. The first is
comparable to the binding energy of a neutron star, the second is
greater than a solar rest mass energy.  Clearly, these values are
problematic for most GRB models.

\subsection{Theory}

If the GRB emission is beamed into some angle, $\theta$, the overall
energy would be lower by a factor $\theta^2 /4$ (the overall event
rate will be larger by the inverse factor).  There has been some
confusion between such ``geometric'' beaming and the relativistic
beaming that arises due to the relativistic motion.  It is worthwhile
to discuss this issue first. The radiation from a source moving with a
Lorentz factor $\gamma$ towards the observer is beamed into a narrow
cone with an opening angle $\gamma^{-1}$.  Typical values during the
GRB and the afterglow are $\gamma \sim 200$ and $\gamma \sim 2-10$
respectively, corresponding to a relativistic beaming of $10^{-2}$rad
and $0.1-0.5$rad. These are the maximal (smallest angle) beaming
possible. However, this is just a lower limit on the beaming angle. If
the source has an opening angle $\theta > \gamma^{-1}$ then the
beaming is determined by $\theta$ and not by $\gamma^{-1}$. Observers
with a viewing angle up to $\theta$ from the center can see the
source.  However, each observer sees a local patch whose size is only
$\gamma^{-1}$.  There are $(\theta \gamma)^2$ such patches.  As all
these patches are causally disconnected, different observers that are
more than $\gamma^{-1}$ apart will observe different emitting regions
and may record a a different time profile and a different spectrum
from the same burst. The causally connected regions grow as $\gamma$
decreases.  Thus, during the afterglow there are fewer and fewer such
regions and when $\gamma^{-1} \sim \theta$ there is only one. We will
return to this issue later, when I discuss the implications of this
phenomenon to the implied energy emission and efficiency of GRBs.

What distinguishes the dynamics of a spherical ejecta from a
nonspherical one? Initially there is little difference. The proper
time required for the matter to reach a radial distance $R$ is
$R/c\gamma$. The maximal sideway expansion is therefore
$R/\gamma$. Hence the angular size of a causally connected region is
$\gamma^{-1}$. As long as $\gamma^{-1} < \theta$ the matter simply
does not have enough time to expand sideways and to know that it is
not a part of a spherical shell \cite{Piran95}. However, once
$\gamma^{-1} \approx \theta$ the matter suddenly ``discovers'' its
nonspherical structure and it begins to expand sideways. As the matter
at the front is constantly shocked to relativistic energies Sari,
Piran and Halpern \cite{SaPH99} expect it to expand with the speed of
light: $\theta \sim \gamma^{-1}$.  Rhoads \cite{Rhoads99} assumes that
this sideway expansion is at the sound speed which results in $\theta
\sim \gamma^{-1}/\sqrt 3$.  This sideway expansion is so rapid that it
dominates completely the radial expansion. For an expansion into a
homogeneous ejecta this 
yields \cite{NP99}\footnote{Rhoads \cite{Rhoads99} 
obtained a different exponent. For
clarity we drive this expression in an Appendix.
This does not
influence, however, the rest of Rhoads' conclusions.} 
$\gamma \propto R^{-3/2}
\exp [-[3/ (2 \gamma_o \theta_{o}) 
( R^{3/2}/ R_{o}^{3/2}-1)]$, which is valid before, during and after 
the transition. Additionally
the radiation is now beamed into a larger cone since $\gamma^{-1} >
\theta_0$.  Both effects reduce the observed emission and will cause a
break in the light curve, roughly by an additional 
factor\footnote{The beaming break  takes place at $\gamma\approx
\theta_0^{-1}$. If the hydrodynamic break is only at $\gamma \approx 
(\sqrt 3\theta_0)^{-1}$ then two successive breaks will take
place \cite{Meszaros98}.}  of $t^{-1}$.  Within the adiabatic
synchrotron model we have new relations between $\alpha$, $\beta$ and
$p$:
\begin{equation}
\alpha_{beam} =\cases { 2 \beta +1 = p &  for $\nu < \nu_c$,
\cr 2 \beta  = p & for $\nu > \nu_c$.}  
\label{beta_jet}
\end{equation}

\subsection{Observations}

\begin{figure}
\centerline{\epsfig{file=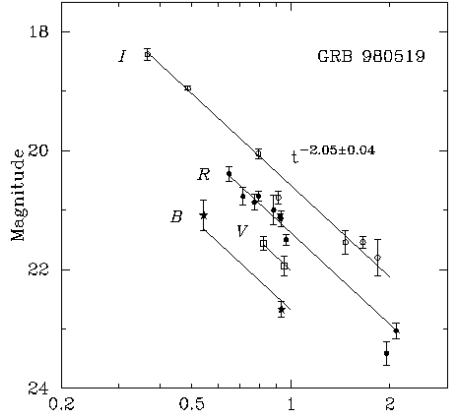,width=6cm}}
\caption{An optical light curve for GRB 980519. The fast decline indicate
a spreading beam \cite{Halpernetal99}. } 
\label{980519}
\end{figure}

The notion of beamed emission was accepted with some difficulty at
first as the first two and up to now the longest and best observed
light curves from GRB 970228 and GRB 970508 show a single power law
decay with no indication for a beaming break. However, other bursts
were different. GRB 980519 was
unique among GRB afterglows with its most rapidly fading $t^{-2.05\pm
0.04}$ in optical as well as in X-rays \cite{Halpernetal99}. The
optical spectrum of this burst shows: $\beta=1.15 \pm 0.$ (see
Fig. \ref{980519}). These values are in perfect agreement with
an expanding beamed emission with $p \approx 2.2$.  As transition is not
seen in the light curve the sideway spreading must have begun before
the first optical observation, namely, less than 8.5 hours after the
burst. Using the detector's time equation (equation \ref{time}) we
find that the corresponding opening angle must have been rather
small: $\theta <0.05$, leading to a beaming factor of 500 or larger!

GRB~980326 was another burst with a rapid decline. Groot et
al. \cite{Groot98} derived a temporal decay slope of $\alpha =2.1\pm
0.13$ and a spectral slope of $\beta =0.66\pm 0.7$ in the optical
band, suggesting once more beaming.  As Groot et al. \cite{Groot98}
note, the large uncertainty in the spectral index allows in this case
also a spherical expansion interpretation (with somewhat unusual
values $p=4.2$ or $p=5.2$).  However, this measured temporal decay was
dependent upon a report of a host galaxy detection at $R=25.5\pm 0.5$,
which was included as a constant term. The detection of a host has
since been determined to be spurious; better data show no constant
component to a limiting magnitude of \cite{Bloometal99} $R=27.3$.  If
the last detection is interpreted as a different
phenomenon \cite{Bloometal99} then the remaining points show a rapid
decline - in agreement with a spreading beam interpretation.

GRB 990123 provided the first direct evidence for a beaming
break \cite{Kulkarni99a}.  The prompt optical flash, which we interpret
to arise from the reverse shock decayed like $t^{-2}$, and disappeared
quickly. The intermediate optical afterglow showed a power law decay
with $t^{-1.1\pm 0.03}$. The decay
 and the spectrum fit well an electron distribution
with $p=2.5.$
This behavior continued from the first late
observation (about 3.5 hours after the burst) until about $2.04\pm
0.46$ days after the burst. Then the optical emission began to decline
faster. The simplest explanation is that we have observed the
transition from a spherical like phase to a sideway expanding
phase. The transition took place at $\sim 2$ days, corresponding to
$\theta _{0}\sim 0.1$. This implies a beaming factor of about $100$
reducing the energy of the burst to $3\times 10^{52}$ ergs.

Finally, just last May, GRB 990510 depicted a beautiful transition
from $t^{-0.82\pm 0.02}$ to $t^{-2.18 \pm 0.05}$ at $1.2\pm 0.08$days
after the burst \cite{Harrisonetal99}. The isotropic energy of this
burst is $2.7 \times 10^{53}$ergs. With a beaming factor of 300 this
becomes a ``modest" $10^{51}$ ergs.

\subsection{On Jets, Bullets or  Flying Pancakes}

Probably because of the analogy with AGNs in which beautiful jets are
observed the terminology jets has been used to describe beamed
emission from GRBs. Unfortunately this terminology is somewhat
misleading. A jet is long and narrow. It corresponds to continuous
activity. The jets seen in AGNs reflect activity on scales of $10^7$
to $10^8$ years. Much longer than the typical time scales of the inner
black holes. Here we observe a transient phenomenon and the length of
the ejecta (in the  direction along its motion) is much shorter.  At first
one may think that the term bullet would be more appropriate.  A
bullet is not so extended in length and it clearly represents a
transient phenomenon.  However, a more detailed consideration reveals
that the perpendicular size of the ejecta is larger than its length.
The angular size is $R\theta$ and the ``length'' is $\Delta \sim c T$.
Typical values are $\Delta \sim 10^{12}$cm (for $T=30$sec) and $\theta
\sim 0.1$. Thus, quite generally, $R\theta> \Delta$.  This is not even
a bullet. This is a pancake flying at relativistic velocity
perpendicular to its flat direction (see Fig. \ref{internal}). One may
wonder if relativistic contraction has played some tricks on us. When
we look at the ejecta in its own rest frame we find that it is longer
by a factor $\gamma$ so initially it is actually a bullet. However,
even at this early phase the ejecta expands sideways proportionaly to
$R$ (unless it is continously collimated) and even in its own frame is
looks like a pancake.

\section{On The Nature Of The ``Inner  Engine''}

We turn now to summarize the evidence concerning the ``inner engine'':
\begin{itemize}
\item {\bf Energy:} The energy,  $\sim 10^{51}-10^{52}$ergs, is 
a significant fraction of the binding energy of a solar mass compact
object.
\item {\bf Internal Shocks:}  To produce internal shocks the ``inner engine''
must produce long and irregular wind.  Single explosions won't work.

\item {\bf Variability} The variability time scale, $\delta t$, suggests a
solar mass compact object. 

\item {\bf Duration:} The duration, $T$, suggests a prolonged activity 
which is much longer than the source's gravitational time scale.

\item{\bf Relativistic Flow:} The central engine must produce 
efficiently a relativistic flow. The 
baryonic load is  less than about $10^{-5}M_\odot$.

\item{\bf Beaming:} The emitted flow is  sometimes highly nonspherical
with an opening angle of few degrees.  Since late afterglow is less
beamed than the GRB this suggest a search for ``orphan'' afterglows
that are not accompanied by GRBs. This also suggest some similarity to
AGNs.

\item{\bf Rate:} The observed rate corresponds approximately 
to one burst per $10^7$years per galaxy. The actual rate may be higher
by a factor of a hundred or so if most GRBs are beamed.  The rate is
still uncertain. Because of the width of the GRB luminosity function
(more than a factor of a hundred) one cannot easily infer the rate of
GRBs from the $\log N/\log S$ distribution.

\item{\bf Host Galaxies:} Host galaxies have been detected for 
most GRBs with optical afterglow. The long standing no-host problem
has disappeared. Most bursts are located at the central regions
indicating that the progenitors have not escaped from their host
galaxies, as would have been the case in long lived binary neutron
stars in dwarf galaxies \cite{NPP92}.

\item{\bf Association with star forming regions:} There is some evidence
that host galaxies of GRBs are star forming
galaxies \cite{HoggFruchter} (see however  \cite{Djorgovski99}).  This
would support short lived progenitors: collapsing massive stars or
short lived\footnote{The life time of the observed galactic neutron
star binaries is $10^8-10^9$years. On the other hand there is a strong
observational bias against detection of short lived neutron star
binaries.} ($10^4-10^5$ years) binary neutron
stars \cite{Tutukov_Yungelson94}.

\item{\bf GRB 980425, SN1999bw and the GRB-SN association:}
The error box of GRB 990425 contains the bright supernova SN1999bw
which is at $z=0.085$ \cite{Galamaetal98b}.  This has lead to a debate
on the association of GRBs with type Ic
supernovae \cite{WangWheeler,Kippenetal98,Grazianietal98}.

\end{itemize}

This evidence, and more specifically, the energy and the time scale
considerations suggest that GRBs are powered by accretion of a massive
$\sim 0.1M_\odot$ accretion disk onto a compact object, most likely a
black hole.  Such a massive accretion disk must form simultaneously
with the black hole, from matter that was slowed down by
centrifugal forces. Thus GRBs signal the formation of black holes.
The gravitational energy of a $0.1M_\odot$ can supply the energy
required for the process, the accretion time would determine the
overall duration while the variability would be determined by the
gravitational time scale of the central object or by the hydrodynamic
time scale of the accretion disk. Both are rather short. A variation
on the theme involves the Blandford-Znajek
effect \cite{Blandford_Znajek77} which extract  the rotational
energy of the black hole - a larger reservoir than the gravitational
energy of the accreting disk. As this effect involves electromagnetic
fields it is likely that it is easier to produce, using this effect, a
clean relativistic flow, possibly Poynting flux.

Several routes  can lead to a black hole - accretion torus system:  
\begin{itemize}
\item{\bf Binary neutron star merger:} Binary neutron star 
mergers \cite{Eichler89} have been considered the canonical
cosmological GRB sources for some time.  These mergers are known to
take place at a rate of one per $\sim 10^6-10^7$ years per
galaxy \cite{NPS91}.  This rate is comparable to the rate of
GRBs \cite{PNS91}. The final outcome of such a
merger \cite{Daviesetal94,RuffetJanka99,Rosswog99} is a $ \sim 2.4
M_\odot$ black hole surrounded by a $0.1-0.2 M_\odot$ thick accretion
disk, which could power the burst \cite{NPP92}.

\item{\bf Neutron star - Black hole merger:} This is a simple variation on the
previous theme. One expect here that the neutron star will be torn
apart by the black hole again leaving a massive disk (possibly
slightly more massive than in the binary neutron star case), which
will power the burst.  If the mass of the black hole is $\sim
10M_\odot$ we expect slightly different time scales and different
behavior between the neutron star binary and this case.  While black
hole - neutron star binaries are expected to be as common as neutron
star - neutron star binaries \cite{NPS91}, or even more
common \cite{BetheBrown98} none was observed so far.

\item{\bf ``Failed Supernova'', Collapsar of Hypernova:} These are all
different name for a collapsing star that produces a GRB.  Both
Woosley's \cite{Woosley93} failed supernova model and
Paczy\'nski's \cite{Pac98} hypernova's model assume that the rapidly 
rotating collapse produces a rotating massive black hole surrounded by
a thick torus that accrets on it.

\item{\bf White dwarf - Neutron star  merger:} A white dwarf
orbiting a neutron star will be pushed inwards via
gravitational radiation emission. If the mass ratio is small it will
become unstable
\footnote{The expected mass ratio in a white dwarf - black hole 
binary is small and we expect these systems to be stable.}  when it
reached the Roche limit and it will dump its mass into an accretion
disk within its hydrodynamics time scale, a few seconds, producing a
solar mass torus surrounding a neutron star. Once accretion from the
disk begins this neutron star will turn into a black hole.  Depending
on the viscosity and on neutrino losses the accretion rate may be high
enough so that the accretion time would be of order few seconds which
will be the duration of the GRB.
\end{itemize}

It is interesting to note that there this sources can be arranged as a
sequence in terms of their maximal time scale with neutron star
binaries the shortest, followed by black hole - neutron star merger, a
white dwarf - neutron star merger and a failed supernova.
The same sequence also arranges the sources from less to more 
baryons surrounding the source.

\section{Some Open Questions}
In a conference called ``Some Open Questions in Astrophysics'' that
took place in the fall of 1995, I \cite{Piran95} have summarized our
understanding of GRBs by the following four open questions: Where?
What? How? and Why?  This was of course an exaggeration. Still it
reflected the ongoing debates at that time.

This has changed drastically. We know now with certainty that GRBs are
cosmological. As for how - the ``Internal-External Shocks,
Relativistic Fireball Model'' is supported very well by all recent
observations. What is producing the GRBs is not determined yet. But
there is a reasonably good case for the ``Accretion onto a Newborn
Black Hole'' scenario, with binary neutron star mergers and Failed
Supernova-Collapsar-Hypernova competing on its origin.

The question, why - or what can we do with these bursts, is still wide
open, even though there have been numerous suggestions on the role
that GRBs could take place, from destroying life\footnote{Quite
unlikely as supernova are much more frequent and hence the expected
fluxes from a ``typical'' nearby supernova would exceed the flux from
a distant galactic GRB.}  to discovering life \cite{Corbert99}. A more
mundane proposals is to explore the high redshift universe and in
particular to measure, using GRBs, the high redshift star formation
rate.  The wide GRB luminosity function failed the idea to use GRBs to
measure cosmological parameters - GRBs are not standard candles. Still
one can obtain, using lensing (or lack of it) some limits on cosmic
parameters \cite{Nemiroff93a}.

The most problematic open question concerning all these models is
the issue of efficiency and the possibility of an ``Energy Crisis'':

\subsection{Energy, Energy Distribution and  Efficiency}
\label{efficiency}

The most basic and fundamental issue is energetics. Is there an energy
crisis? The implied isotropic energy can be as high as $4 \times
10^{54}$ergs.  Even with beaming the energy is reduced to a gigantic
$10^{52}$ergs.  Internal shocks can convert under reasonable
conditions $\sim 10\%$ of the kinetic energy to thermal
energy \cite{MMM95,KPS97,Daigne98,Kumar99}. External shocks are even less
efficient.  Then there is the question of radiative
efficiency \cite{SaNP96,SaP97b} and of additional energy
losses \cite{Kumar99}. Even before all that one should wonder what was
efficiency of the source in accelerating the relativistic flow?  We
are most likely faced with an energy crisis.  This had led some to
resort to more powerful central engines, namely more massive compact
object such as $10 M_\odot$ black hole powered by rotation or by mode
massive accretion disk. But there is a limit to this route as some
bursts show submillisecond variability that would pose an upper limit
on the central mass.

A related issue is the question of the energy distribution between the
GRB and its afterglow. According to the internal-external shocks model
a comparable amount of energy should be released in the GRB and in the
early afterglow. Actually the early afterglow should be more powerful
than the GRB by a factor of a few. However, early afterglow
observations show the opposite. The prompt X-ray emission (which could
be interpreted as early afterglow) is at most 10\% of the GRB energy.
The impressive optical flash from the reverse shock of GRB 990123 was
only 1\% of the gamma-ray energy. Has anything gone wrong?

There is one simple answer to both question: even after the beaming
correction we are over estimating the GRB energy. There are two
possible reasons for that: First, the GRB luminosity might be
dominated by hot spots \cite{KumarPiran99}, whose angular size
$\gamma^{-1}$ is much smaller that the overall geometrical beaming
$\theta_0$.  During the afterglow these regions grow and their
emission is spread over larger angular regions. This effect could also
explain the wide GRB luminosity function. The observed sample of GRBs
with afterglows is biased towards strong bursts and hence towards
cases in which such a hot spot have been observed.  A second
phenomenon that has similar effect is spreading of the beam from the
internal shocks phase to the early afterglow phase \cite{NP99}. Here we
expect expansion only by a factor of a few, but this could be
sufficient to  explain the discrepancy between the GRB and the
early afterglow luminosities.

\subsection{Acceleration, Shocks and  Microphysics}

A second set of theoretical open questions involve unknown
microphysics: How does the ``inner engine'' accelerate the ejecta  to
relativistic velocities?  How do the collisionless shocks arise within
the emitting regions  ?
How do these shocks accelerate particles and
enhance the magnetic fields(see however \cite{Medvedev_Loeb}) ? 
These  questions are wide
open.  They could be an extremely challenging tasks for ambitious PhD
projects, with not too good chances of success. Should we fail this
model because of this ignorance? Here, one can just look around and
realize that related  question have not been answered in
other, much better studied, astronomical source: AGNs, pulsars and our
own solar corona to name a few.

\subsection{An Observational Wish List}
Because of the accidental nature of GRBs, when discussing
observational issues one can only state a wish list.  We hope that new
bursts will provide answers to this questions, even though some
observers seem to know better how to make a wish list materialize.

The first group of questions deals directly with the sources.  Answer
to these questions could resolve, once for all, the mystery of GRBs.
\begin{itemize}
\item Is there a   GRB-SN association? 
\item What is the relation between  GRBs and star forming regions?
\item Are there  short lived binary neutron star systems and/or
black hole-neutron star binaries?
\end{itemize}

The second group of questions deals with the physical processes.  Here
detailed multiwavelenght light curves of the prompt emission and the
early afterglow could provide invaluable information on the extreme
conditions that take place in these regions. The next generation GRB
detectors with precise position capabilities and rapid response
systems promise that these wishes will be fulfilled in the not to
distant future. Among these the question whether afterglow is generic
and arises also for short or for less intense bursts is an important
and crucial issue.

\section{Epilogue} 

Ten years after the neutron star merger paper \cite{Eichler89}, these
sources are amongst the best candidates for GRB sources. The major
problems facing them are the ``energy crisis" that faces most compact
GRB models and the issue of association with star forming regions,
which is not expected for long lived neutron star binaries. In spite
of those problems even today neutron star mergers are still the only
sources that can produce the enormous amounts of energy involved and
are based on an independently observed phenomenon that is known to
take place at a comparable rate \cite{PNS91}.  Ironically, while the
specific mechanism (pair creation via $\nu \bar \nu$ annihilation)
discussed in this early paper \cite{Eichler89} is most likely invalid
in the context of neutron star mergers, it has been recently suggested
that it might be important in the context of the competing failed
supernovae-collapstar model \cite{MacFadyenWoosley98}.

I have stressed repeatedly here and elsewhere that the GRBs' ``inner
engine'' is hidden.  As such we cannot distinguish directly between
different GRB models that potentially go via the same route and
produce the a black hole - accretion torus system believed to be
capable of powering a GRB. At this stage I can only turn, once more,
to my wish list and add one final wish - the simultaneous detection of
a GRB and a chirping gravitational radiation signal characterizing a
neutron star merger. Such a coincidence will clearly enhance the
significance of the detection of the gravitational radiation
\cite{Kochaneck_Piran}. It will also verify this merger model and will
resolve the GRB enigma.

\section{Acknowledgements}
I thank Re'em Sari for a wonderful and productive collaboration and
J. Granot, J. Katz, S. Kobayashi, P. Kumar, R. Narayan and
F. K. Thielemann, for many helpful discussions.  This research was
supported by the US-Israel BSF grant 95-328, by a grant from the
Israeli Space Agency and by NASA grant NAG5-3516.  I thanks Columbia
University, NYU and Basel University for their hospitality while this
research was done.

\begin{appendix}%

\bigskip
\noindent {\bf Appendix}
\bigskip

Following Narayan and Piran \cite{NP99}  I derive here the exponential
dependence of the Lorentz factor, $\gamma$, on $R$ in an sideway
expanding relativistic beam.  $R$ and $R_\perp$ are along and
perpendicular to the jet's axis in the observer frame.  We write the
adiabatic energy equation (\ref{adiabatic}) as: 
$$
\gamma =\gamma_o {R_{\perp o} R_{o}^{1/2} \over R_\perp R^{1/2}}.
\eqno{(A1)}
$$
The sideway propagation equation is:
$$
{dR_\perp \over d R} = {c_x \over \gamma} + {R_\perp \over R} = 
{c_x R_\perp R^{1/2} \over \gamma_o R_{\perp o} R_{o}^{1/2} }
+ {R_\perp \over R},
\eqno{(A2)}
$$ where $c_x$ is either the speed of light (1) or the speed of sound
$(1/\sqrt 3)$.   Integration yields: 
$$
{R_\perp\over R_{\perp o}} = {R \over R_{o} }
\exp \bigg[{3 \over 2} {c_x \over \gamma_o \theta_{o}} 
( {R^{3/2} \over R_{o}^{3/2}}-1)\bigg].
\eqno{(A3)}
$$
Finally, we  substitute A3 in A1 to obtain:
$$
\gamma =\gamma_o ({R_{0}\over R })^{3/2} 
\exp \bigg[-{3 \over 2} {c_x \over \gamma_o \theta_{o}} 
( {R^{3/2} \over R_{o}^{3/2}}-1)\bigg].
\eqno{(A4)}
$$
\end{appendix}

\end{document}